\documentstyle[aps,prl]{revtex}

\begin{document}
\draft
\title{Proton Fraction in Neutron Stars}
\author{Feng-Shou Zhang$^{1,2,4}$, Lie-Wen Chen$^{1,2,3}$}
\address{$^1$Center of Theoretical Nuclear Physics, National Laboratory of Heavy Ion\\
Accelerator, Lanzhou 730000, China}
\address{$^2$Institute of Modern Physics, Chinese Academy of Sciences, Lanzhou
730000, China}
\address{$^3$Department of Applied Physics, Shanghai Jiao Tong University, Shanghai
200030, China}
\address{$^4$CCAST (World Laboratory), P.O. Box 8730, Beijing 100080, China}
\maketitle

\begin{abstract}
The proton fraction in ${\sl \beta }$-stable neutron stars is investigated
within the framework of the Skyrme-Hartree-Fock theory using the extended
Skyrme effective interaction for the first time. The calculated results show
that the proton fraction disappears at high density, which implies that the
pure neutron matter may exist in the interior of neutron stars. The
incompressibility of the nuclear equation of state is shown to be more
important to determine the proton fraction. Meanwhile, it is indicated that
the addition of muons in neutron stars will change the proton fraction. It
is also found that the higher-order terms of the nuclear symmetry energy
have obvious effects on the proton fraction and the parabolic law of the
nuclear symmetry energy is not enough to determine the proton fraction.
\end{abstract}


The neutron star serves as a cosmic laboratory to study the nuclear equation
of state (EOS) at high density. It has been known that the mass, radius, and
density profile of neutron star are determined by the EOS of asymmetric
nuclear matter, and the chemical composition, particularly the proton
fraction is determined by the nuclear symmetry energy$^1$. Many theoretical
methods, such as the Brueckner approach\ $^2$, variational many body (VMB)
theory$^3$, the relativistic mean field (RMF) theory$^4$, and
non-relativistic many body theory$^5$, have been used to study the proton
fraction in neutron stars since the proton fraction is an important physical
quantity to determine the evolution of neutron stars, such as the cooling
rate, neutrino flux, and so on$^{6-9}$. However, little is known about the
properties of the nuclear symmetry energy at high density$^{10}$ with the
proton fraction in neutron stars remaining undetermined.

In our previous work$^{11,12}$, a density, temperature and isospin dependent
nuclear EOS has been deduced within the framework of the Skyrme-Hartree-Fock
theory using the so-called extended Skyrme effective interaction$^{13}$. For
asymmetric nuclear matter with density $\rho $ and neutron excess $\delta
=(\rho _n-\rho _p)/\rho $ , the energy per nucleon at temperature $T$ can be
written as

\begin{eqnarray}
\varepsilon \left( \rho ,T,\delta \right) &=&\frac 12T\left[ \frac{%
C_{3/2}(\mu _{\tau _\alpha })}{C_{1/2}(\mu _{\tau _\alpha })}(1+\delta
)^{5/3}+\frac{C_{3/2}(\mu _{\tau _\alpha })}{C_{1/2}(\mu _{\tau _\alpha })}%
(1-\delta )^{5/3}\right] +  \nonumber \\
&&\ \ \frac 14\left[ a_{\tau _\alpha }^{(1)}\left( 1+\delta \right)
+a_{-\tau _\alpha }^{(1)}(1-\delta )\right] \rho +  \nonumber \\
&&\ \ \frac 14\left[ a_{\tau _\alpha }^{(2)}\left( 1+\delta \right)
+a_{-\tau _\alpha }^{(2)}(1-\delta )\right] \rho ^{\gamma +1}+  \nonumber \\
&&\ \ \frac 14\left[ a_{\tau _\alpha }^{(3)}\left( 1+\delta \right)
+a_{-\tau _\alpha }^{(3)}(1-\delta )\right] \rho ^{5/3}+  \nonumber \\
&&\ \ \frac 14\left[ a_{\tau _\alpha }^{(4)}\left( 1+\delta \right)
+a_{-\tau _\alpha }^{(4)}(1-\delta )\right] \rho ^{\gamma +5/3},
\end{eqnarray}
with

\[
a_{\tau _\alpha }^{(1)}=\frac 14t_0\left[ 3\mp \left( 2x_0+1\right) \delta
\right] , 
\]

\[
a_{\tau _\alpha }^{(2)}=\frac 1{24}t_3\left[ 3\mp \left( 2x_3+1\right)
\delta \right] , 
\]

\begin{eqnarray*}
a_{\tau _\alpha }^{(3)} &=&\frac 1{16\pi ^2}\left[ t_1\left( 1-x_1\right)
+3t_2(1+x_2)\right] \left( 1\pm \delta \right) ^{5/3}\left( \frac{2\sqrt{\pi 
}}\lambda \right) ^5C_{3/2}\left( \mu _{\tau _\alpha }\right) \\
&&+\frac 1{8\pi ^2}\left[ t_1(1+\frac{x_1}2)+t_2(1+\frac{x_2}2)\right]
\left( 1\mp \delta \right) ^{5/3}\left( \frac{2\sqrt{\pi }}\lambda \right)
^5C_{3/2}\left( \mu _{_{-}\tau _\alpha }\right) ,
\end{eqnarray*}

\begin{eqnarray*}
a_{\tau _\alpha }^{(4)} &=&\frac 1{16\pi ^2}\left[ t_4\left( 1-x_4\right)
+3t_5(1+x_5)\right] \left( 1\pm \delta \right) ^{5/3}\left( \frac{2\sqrt{\pi 
}}\lambda \right) ^5C_{3/2}\left( \mu _{\tau _\alpha }\right) \\
&&\ +\frac 1{8\pi ^2}\left[ t_4\left( 1+\frac{x_4}2\right) +t_5(1+\frac{x_5}2%
)\right] \left( 1\mp \delta \right) ^{5/3}\left( \frac{2\sqrt{\pi }}\lambda
\right) ^5C_{3/2}\left( \mu _{-\tau _\alpha }\right) .
\end{eqnarray*}
where $\tau _\alpha $ and $-\tau _\alpha $ represent neutron and proton,
respectively. Every equation is split into two equations, which are for the
neutron and proton, respectively. The $t_0-t_5$ and $x_0-x_5$ are Skyrme
potential parameters, $\lambda $ and $C_l(\mu _{\tau _\alpha })$ represent
the thermal wave length and Fermi-Dirac integral.The Taylor expansion of Eq.
(1) with respect to $\delta $ at temperature $T=0$ (hereafter,we define $%
\varepsilon (\rho ,\delta )=\varepsilon (\rho ,\delta ,T=0)$) is as follows 
\begin{equation}
\varepsilon (\rho ,\delta )=\varepsilon (\rho ,\delta =0)+C(\rho )\delta
^2+D(\rho )\delta ^4+O(\delta ^6),
\end{equation}
where $C$ is the so-called coefficient of symmetry energy strength.

The chemical composition of the neutron star is determined by the
requirement of charge neutrality and equilibrium with respect to the weak
interaction ( $\beta $-stable matter ). From Eq. (1) or (2) we can calculate
the proton fraction, $x_p=(1-\delta )/2$, for $\beta $-stable matter as
found in the interior of neutron stars. Equilibrium for the reactions $%
n\rightarrow p+e^{-}+\overline{\nu }_e$ and $p+e^{-}\rightarrow n+\nu _e$
requires that 
\begin{equation}
\mu _n=\mu _p+\mu _e,
\end{equation}
where $\mu _i=\partial \varepsilon /\partial N_i$ ($i=n$, $p$, $e$ and $\mu $%
) is the chemical potential. Consequently, one obtains 
\begin{equation}
\mu _n-\mu _p=2\frac{\partial \varepsilon }{\partial \delta }.
\end{equation}
For relativistic degenerate electrons, 
\begin{equation}
\mu _e=(m_e^2+k_{Fe}^2)^{1/2}=\left[ m_e^2+(3\pi ^2\rho x_e)^{2/3}\right]
^{1/2},
\end{equation}
with $\hbar =c=1$ and $x_p=x_e$ because of charge neutrality. From Eqs. (1),
(3)-(5), it is easy to find numerically the proton fraction as a function of
density $\rho $.

Just above nuclear matter density at which $\mu _e$ exceeds the muon mass $%
m_\mu $, the reactions $e^{-}\rightarrow \mu ^{-}+\nu _e+\overline{\nu }_\mu 
$, $p+\mu ^{-}\rightarrow n+\nu _\mu $ and $n\rightarrow p+\mu ^{-}+%
\overline{\nu }_\mu $ may be energetically allowed such that both electrons
and muons are present in $\beta $-stable matter. This alters the $\beta $%
-stability condition to 
\begin{equation}
\mu _n-\mu _p=\mu _e=\mu _\mu =\left[ m_\mu ^2+(3\pi ^2\rho x_\mu
)^{2/3}\right] ^{1/2},
\end{equation}
with $x_p=x_e+x_\mu $

The bulk properties of symmetry nuclear matter at saturation density, such
as the density $\rho _0$, the binding energy per nucleon $\varepsilon _0$,
the incompressibility $K$, the effective mass $m^{*}/m$, and the symmetry
energy are summarized in Table 1 for the three parameter sets SII, SIII, and
SkI5. From Table 1, one can see that the difference for incompressibility
between SII and SkI5 is very large, about 85 MeV at saturation density,
while they almost have the same $m^{*}/m$=0.58. Meanwhile, the difference
for $m^{*}/m$ is very large between SIII and SII, about 0.18 at saturation
density, while the difference for incompressibility between them is
relatively small, about 14 MeV. In calculations, we note that the binding
energy per nucleon $\varepsilon $ and the nuclear symmetry energy $C$
exhibit completely different behaviors with the increase of the baryon
density for different Skyrme parameter sets. These differences come from the
difference of the nuclear EOSs. Therefore, one may investigate the proton
fraction for different nuclear EOSs.

Under the parabolic approximation (up to the -$\delta ^2$ term in Eq. (2)),
the chemical compositions of neutron stars in both the electrons only and
electrons plus muons cases as a function of the baryon density are shown in
Figs. 1 (a)-1 (c) using the Skyrme parameter sets SII, SIII, and SkI5,
respectively. In order to ensure validity of the proton fraction above, we
show the squared ratio of sound speed to light speed, ($s/c$)$^2=\partial
P/\partial E$ (where $E$ is the total energy density), in Fig. 1 (d). The
causality condition requires ($s/c$)$^2\leq 1$. One notes that for SII and
SkI5, the squared ratio ($s/c$)$^2$ will make the causality condition fail
to be satisfied at baryon density larger than 1.35 fm$^{-3}$ and 1.1 fm$%
^{-3} $, respectively, about 8 times larger than the saturation density for
symmetry nuclear matter. For SIII, up to baryon density 1.5 fm$^{-3}$, the
causality condition can be still satisfied. From Figs. 1 (a)-1 (c), one can
find that the addition of muons increases significantly the proton fraction
over a broad range of density. In case of SIII, the muon fraction is very
small and muons only exist at about 0.16 fm$^{-3}$, which is due to the fact
that $\mu _e$ is too small to exceed the muon mass $m_\mu $. The value of
the proton fraction is an important ingredient to model the thermal
evolution of a neutron stars. In fact, if the proton fraction in the
interior of star is above a critical value $x^{Urca}$, the so-called direct
Urca processes $n\rightarrow p+e^{-}+\overline{\nu }_e$ and $%
p+e^{-}\rightarrow n+\nu _e$ can occur$^{14}$. If they occur, neutrino
emission and neutron star cooling rate are increased by a large factor
compared to the standard cooling scenario$^{14}$. The critical proton
fraction has been estimated to be in the range of $x^{Urca}=11-15\%$. It is
indicated from Fig. 1 that the proton fraction for SII and SIII cannot reach
this range. However, the proton fraction reaches $x^{Urca}=11\%$ at baryon
density 0.2 fm $^{-3}$ for SkI5. This means that one can get a reasonable
proton fraction for a softer nuclear EOS, such as $K=255$ MeV for case of
SkI5. While for stiffer ones, such as SII ($K=341$ MeV) and SIII ($K=355$
MeV), one might not get reasonable proton fraction although they may have a
very different effective mass ($m^{*}/m=$ 0.580 and 0.763 for SII and SIII,
respectively). It seems that the incompressibility is more important to
determine the proton fraction.

One of the interesting features of Fig. 1 is the disappearance of the proton
fraction at high density, namely, $\rho =0.48$ and $\rho =0.32$ fm$^{-3}$
for SII and SIII, respectively. This phenomenon may be due primarily to the
greater short-range repulsion in isospin singlet nucleon pairs compared to
isospin triplet nucleon pairs$^{15}$. At high density this short-range
repulsion must dominate and pure neutron matter is favored. At intermediate
densities the strong isospin singlet tensor potential and correlation serve
to keep the isospin singlet nucleon pairs, and thus symmetry nuclear matter,
more attractive than pure neutron matter. These features imply that the pure
neutron matter may exist in the interior of neutron stars where the density
is high enough. This phenomenon has also been observed in the calculations
of the VMB theory. For parameter set SkI5, the disappearance of the proton
fraction is not found up to 1.5 fm$^{-3}$.

Many theoretical studies have shown that the empirical parabolic law of the
single particle binding energy is well satisfied within a broad range of
density and neutron excess $\delta $, namely, in Eq. (2) the $\delta ^2$
term is significant, but the higher-order terms are very small. However, it
is still unknown how the higher-order terms affect the proton fraction at
high density. Fig. 2 is the same as Fig. 1 except for including all the
higher-order terms. From the comparison between Figs. 1 and 2, one can find
that the addition of the high-order terms increases significantly the
fraction of different constituents, $p$, $e^{-}$ and $\mu ^{-}$, over a
broad range of density for parameter sets SII and SIII. It implies that the
higher-order terms of the nuclear symmetry energy have obvious effects on
the chemical composition and the parabolic approximation of the nuclear
symmetry energy may result in large deviation.

The authors would like to thank Professor Peng Qiu-He for interesting
discussions. This work was supported by the National Natural Science
Foundation of China under Grant Nos. 19875068 and 19847002, the Major State
Basic Research Development Program under Contract No. G2000077407, and the
Foundation of the Chinese Academy of Sciences.

\section*{Table Captions}

\begin{description}
\item[Table I]  Bulk properties of symmetry nuclear matter at saturation
point.

\begin{center}
{\small 
\begin{tabular}{cccc}
\hline\hline
Parameter & SII & SIII & SkI5 \\ \hline
$\rho _0$ (fm$^{-3}$) & 0.1484 & 0.1453 & 0.1558 \\ 
$\varepsilon _0$ (MeV/nucleon) & -15.99 & -15.85 & -15.85 \\ 
$K$ (MeV) & 341.38 & 355.35 & 255.78 \\ 
$m^{*}/m$ & 0.580 & 0.763 & 0.579 \\ 
$C$ (MeV) & 34.26 & 28.25 & 36.75 \\ \hline\hline
\end{tabular}
}
\end{center}
\end{description}

\section*{Figure captions}

\begin{description}
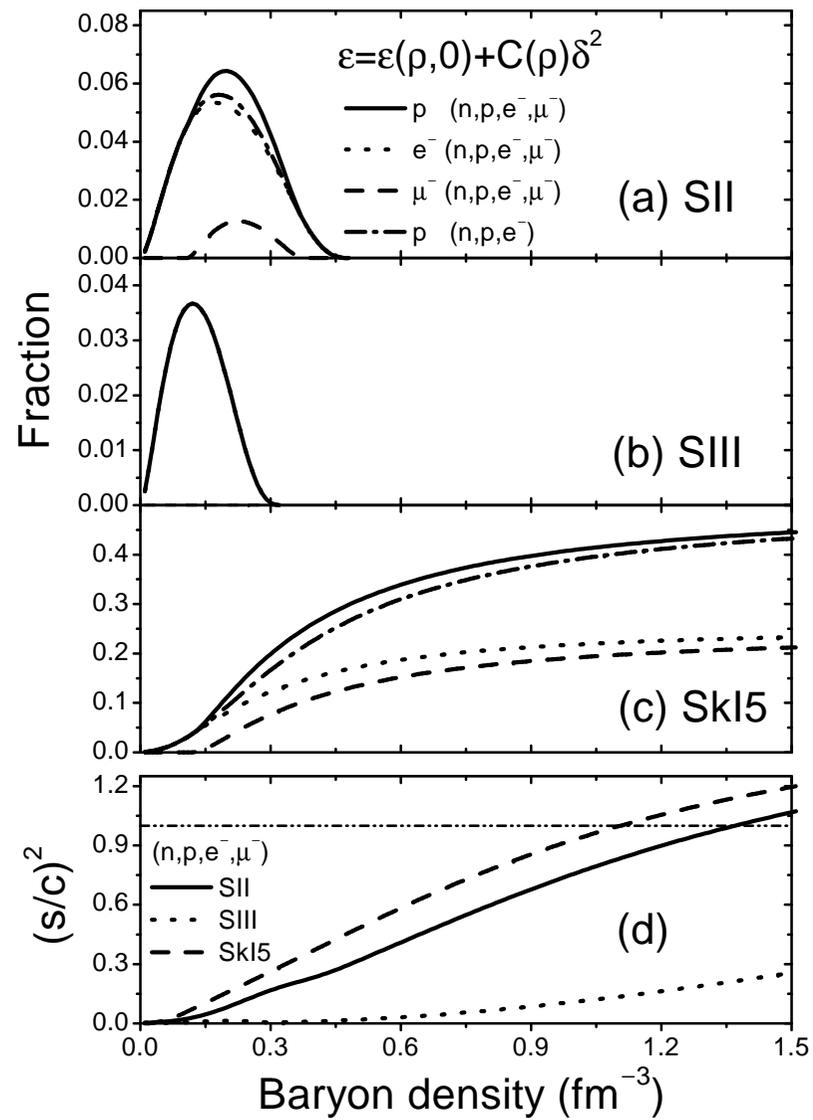
\item[Fig. 1]  With parabolic approximation, the compositions of neutron
stars in both the electrons only and electrons plus muons cases as a
function of the baryon density using the Skyrme parameter sets SII (a), SIII
(b), SkI5 (c). Meanwhile, the squared ratio of sound speed to the light
speed ($s/c$)$^2$ as a function of the baryon density for these three
parameter sets is displayed in (d) in electrons plus muons case. In (b), the
muon fraction is very small so that the solid line, dotted line and
dash-dotted line coincide with each other.

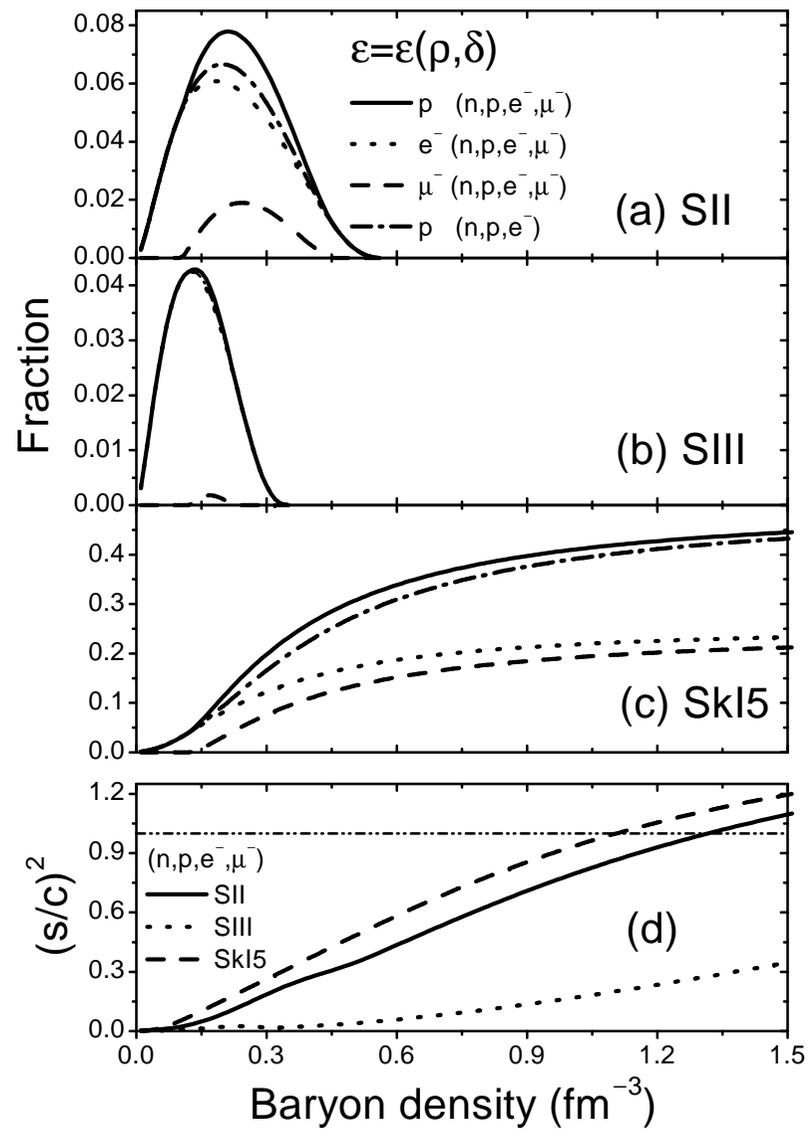
\item[Fig. 2]  Same as Fig. 1 but including all higher-order terms instead
of parabolic approximation for nuclear symmetry energy.
\end{description}


\begin{references}
\bibitem{Physrep97}  M. Prakash et al., Phys. Rep. {\bf 280}, (1997) 1.

\bibitem{brueckner2}  B. Ter Haar and R. Malfliet, Phys. Rev. C 50, (1994)
31.

\bibitem{vmb3}  R. B. Wiringa et al., Phys. Rev. C 38, (1988) 1010.

\bibitem{rmf4}  B. D. Serot et al., Adv. Nucl. Phys. 16, (1986) 1.

\bibitem{nre3}  D. T. Khoa et al., Nucl. Phys. A602, (1996) 98.

\bibitem{SumiAA95}  K. Sumiyoshi et al., Astron. Astrophys. 303, (1995) 475.

\bibitem{cpl1}  S. Zhang, T. P. Li, and M. Wu, Chin. Phys. Lett. 15, (1998)
74.

\bibitem{cpl2}  Y. F. Huang, Z. G. Dai, and T. Lu, Chin. Phys. Lett. 15,
(1998) 775.

\bibitem{cpl3}  K. X. Xu, G. J. Qiao, Chin. Phys. Lett. 16, (1999) 778.

\bibitem{LbaPRL97}  B. A. Li et al., Phys. Rev. Lett. 79 (1997) 1644.

\bibitem{zfs96}  F. S. Zhang, Z. Phys. A356, (1996) 163.

\bibitem{zfs98}  F. S. Zhang and L. X. Ge, Nuclear Multifragmentation,
Science Press, Beijing, 1998, p242 (in Chinese).

\bibitem{Zhuo85}  Y. Z. Zhuo et al., Adv. Sci. in China, Phys. 1, (1985) 231.

\bibitem{LattimerPRL91}  J. Lattimer et al., Phys. Rev. Lett.{\bf \ }66,
(1991) 2701.

\bibitem{PandharipandPLB72}  V. R. Pandharipande et al., Phys. Lett. B39,
(1972) 608.
\end{references}
\end{document}